\documentclass[aps,prd,10pt,onecolumn,nofootinbib,superscriptaddress,floatfix]{revtex4-2}

\usepackage[T1]{fontenc}
\usepackage{lmodern}
\usepackage[utf8]{inputenc}
\ifdefined\pdfglyphtounicode
  \input{glyphtounicode}
\fi
\usepackage{amsmath,amssymb,mathtools}
\usepackage{bm}
\usepackage{graphicx}
\usepackage{placeins}
\usepackage{hyperref}

\hypersetup{
  colorlinks=true,
  linkcolor=blue,
  citecolor=blue,
  urlcolor=blue
}

\begin{document}
\raggedbottom

\title{Can a Wormhole Induce a Monopole?}

\author{Sergei Slobodov}
\email{sergei@slobodov.com}
\noaffiliation

\date{\today}

\begin{abstract}
Could a star orbiting one mouth of a wormhole make the mass seen near the other
rise and fall? We show that the periodic mass change proposed for searches
around Sgr~A* does not occur. The predicted signal comes from treating static
solutions with different masses as successive states of the same orbiting
system. In the Maxwell analogue, charge measured around a mouth changes only
when current crosses the measuring surface. Similarly, in gravity, source
motion cannot change the observed Schwarzschild mass at first order.
Furthermore, the Bondi mass of a vacuum observer end cannot increase even
beyond first order, though gravitational radiation can reduce it. The same
observer-side dynamics can be reproduced with both mouths in one exterior
at sufficiently large separation.
\end{abstract}

\maketitle

\section{Introduction}
\label{sec:introduction}

Seen from far away, the mouth of a traversable wormhole looks like any other
compact mass. Matter near the far mouth could perturb stellar orbits near ours,
allowing us to distinguish the wormhole from an ordinary compact mass.
Dai and Stojkovic (DS)
proposed a concrete version of this idea~\cite{DaiStojkovic2019Observing}.  
In their model, two Schwarzschild exteriors
are joined at a throat of radius $R$, and a mass $\mu$ is placed at radius $A$ in
one of them.  Their static calculation gives a mass shift $b=\mu R/A$ seen
from the other exterior.  Letting $A$ follow an eccentric orbit then makes
this mass rise and fall, producing a periodic $1/r^2$ acceleration of nearby
stars. Proposed searches around Sgr~A* use this signal
\cite{SimonettiEtAl2021SensitiveSearches,MoiseevSazhina2025ObservationalSearch}.

The proposed signal requires the inferred mass to change as the source moves.
In the Maxwell analogue, charge conservation rules out such a response unless
current crosses the measuring surface. Einstein's equations likewise fix the
observer-side mass monopole at first order about Schwarzschild. Beyond first order, the
observer-end Bondi mass cannot increase. Each result requires different assumptions:

\begin{itemize}
\item The measuring surface around the observer's mouth lies in vacuum, so
  no matter crosses it.  This is the situation DS consider: the source stays
  on the other side.
\item For the first-order result, the background near the observer is
  Schwarzschild, as in the DS model, and the source mass is small,
  $\mu/M\ll1$.
\item For the result beyond first order, the observer's end is assumed
  asymptotically flat and vacuum near future null infinity, so that its Bondi
  mass is well defined. For one-ended wormholes, the result applies in the
  large-separation limit described in Sec.~\ref{sec:nonlinear}.
\end{itemize}

The DS prescription gives a family of source-position-dependent static
monopoles. It is natural to impose the same asymptotic normalization at both
ends: equal potentials in the Maxwell calculation and equal clock rates in
gravity. But maintaining that choice as the source moves produces a
time-varying monopole, violating charge conservation in Maxwell theory and
the vacuum Einstein equations at first order in gravity. The proposed
cross-throat mass monopole induction effect is therefore not present.
Time-varying higher multipoles, however, are allowed.

Sections~\ref{sec:maxwell}--\ref{sec:nonlinear} treat the charge, the
first-order mass and the Bondi mass in turn. We also estimate the allowed
quadrupolar influence on the observer through the wormhole.

We use $G=c=1$ and $r_g=2M$, where $M$ is the background Schwarzschild mass.

\subsubsection*{Relation to earlier work}

For a spherically symmetric configuration, Krasnikov pointed out that
Birkhoff's theorem already excludes the DS signal
\cite{Krasnikov2020CommentObserving}.  Our first-order result extends this to
nonspherical source motion, and the Bondi argument extends it beyond first
order. The need to fix the charge sector also appears in Linet's correction
to Copson's black-hole electrostatic potential
\cite{Copson1928,Linet1976} (Sec.~\ref{sec:maxwell}).

\section{Conservation of electric monopole charge}
\label{sec:maxwell}

Here we show that there is no cross-throat monopole induction in the Maxwell case. 
The electric monopole is the electric flux through
a closed surface $S$ surrounding the observer's mouth.  In terms of the
electromagnetic field strength $F$ and its Hodge dual $\star F$, the measured
charge is
\begin{equation}
 Q[S]=\frac{1}{4\pi}\int_S\star F.
 \label{eq:maxwell-charge}
\end{equation}
A family of such surfaces sweeps out a timelike worldtube $\mathcal T$
with boundary $S_{t_2}-S_{t_1}$.  Maxwell's equation is
$d\star F=4\pi\star j$, where $j$ is the electric current one-form.
Integrating over the worldtube and applying Stokes' theorem gives
\begin{equation}
 Q[S_{t_2}]-Q[S_{t_1}]=\int_{\mathcal T}\star j.
 \label{eq:maxwell-balance}
\end{equation}
Without current across $\mathcal T$, source motion cannot change $Q[S]$.
The proof uses the worldtube, not an enclosed spatial volume, and requires
neither a compact interior nor an asymptotic end.  It remains exact on an
evolving metric; electromagnetic radiation carries energy, not charge.
Frolov and Novikov established this flux balance
for wormholes \cite{FrolovNovikov1990PhysicalEffects}, and Krasnikov solved the fixed-flux problem for a point charge
\cite{Krasnikov2008ElectrostaticInteraction}.

\begin{figure}[!htbp]
  \centering
  \includegraphics[width=0.82\textwidth]{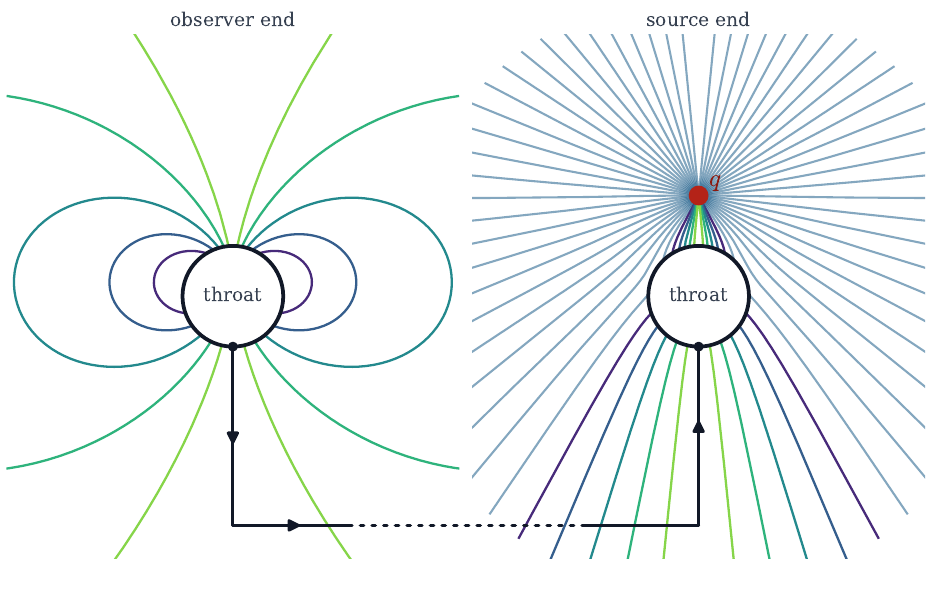}
  \caption{The zero-flux member of Eq.~\eqref{eq:ds-multipoles}: $B_0=0$,
  requiring $c_2-c_1=q/A$.  The black circles represent the same throat sphere
  in the two exterior charts.  Field lines cross the throat but return through
  another angular band, giving zero net flux.  The
  observer-side field is nonzero and contains only $\ell\ge1$ terms.
  For a widely separated one-ended handle, the arrowed path through the throat
  and the dotted return through the exterior form a loop.
  The plotting code was prepared with AI assistance; the author verified the figure.}
  \label{fig:fixed-sector-field}
\end{figure}

Charge conservation fixes the monopole, but higher multipoles can respond
to source motion. Consider
the static family in the DS geometry: two flat exteriors joined at $r=R$,
with charge $q$ at $r_1=A$ in one exterior. We allow the asymptotic potentials
$c_1,c_2$ at the two ends to depend on $A$.
Matching the potential and its normal derivative gives
\begin{equation}
 V_2=c_2+\sum_{\ell\ge0}\frac{B_\ell}{r_2^{\ell+1}}
 P_\ell(\cos\theta_2),\qquad
 B_{\ell\ge1}=q\,\frac{2\ell+1}{2(\ell+1)}
 \frac{R^{2\ell+1}}{A^{\ell+1}},\qquad
 B_0=\frac{qR}{2A}+\frac{R}{2}(c_1-c_2).
 \label{eq:ds-multipoles}
\end{equation}
Here $P_\ell$ is a Legendre polynomial and the observer-side charge is
$Q_2=B_0$.  In the quasistatic description, holding $Q_2$ fixed requires
$c_1-c_2$ to adjust as $A$ changes, while the higher multipoles respond.
Figure~\ref{fig:fixed-sector-field} shows the case $Q_2=0$: the
observer-side field is nonzero but contains only higher multipoles.

DS instead set $c_1=c_2$, obtaining $Q_2=qR/(2A)$
\cite{DaiStojkovic2019Observing}.
Imposing this condition anew at each $A$ selects different charges, not
successive states of a no-current evolution.  The DS relation $Q_1+Q_2=q$,
where $Q_1$ and $Q_2$ are the source-side and observer-side charges, respectively, 
conserves only the sum, whereas Eq.~\eqref{eq:maxwell-balance}
conserves each charge separately when
neither measuring worldtube is crossed.

For a one-ended handle, the freedom corresponding to the cross-end potential difference
is circulation of $\mathbf E$
around a loop through the throat and back through the exterior
\cite{FrolovNovikov1990PhysicalEffects}.
Requiring a single-valued static potential sets this circulation to zero,
just as requiring $c_1=c_2$ for all $A$ fixes the two-ended potential difference, breaking compatibility with
time evolution as $A$ changes.
Preserving zero initial flux instead gives a time-varying circulation magnitude of about $|q|/A$
when the mouths' exterior separation $D$ is much larger than $A$.

Selecting a source-free monopole field to preserve the prescribed charge is
familiar from black-hole electrostatics.  Copson's potential for a static
point charge $q$ at radius $A$ outside a
Schwarzschild black hole \cite{Copson1928} leaves the hole with the charge
$-qM/A$.  Linet removed it with the source-free monopole that keeps the hole
neutral \cite{Linet1976}, and Smith and Will computed the self-force in this
sector \cite{SmithWill1980}.
In Eq.~\eqref{eq:ds-multipoles}, the source-free term
$\tfrac{R}{2}(c_1-c_2)$ plays the role of Linet's monopole.

For the Ellis wormhole, Beskin et al.\ fix the source-free monopole by
preserving an initially zero observer-end flux during quasistatic source
motion \cite[Secs.~4,6]{BeskinEtAl2011ElectricMagnetic}.  Boisseau and Linet
explicitly identify the same zero-flux sector
\cite{BoisseauLinet2013Electrostatics}.  Equation~\eqref{eq:maxwell-balance}
gives this constraint without a quasistatic approximation.

Like DS, Rub\'{\i}n de Celis et al.\ impose equal
asymptotic potentials.  They propose distinguishing a wormhole from a neutral
black hole by its attractive far-field self-force
\cite[Eq.~(3.19)]{RubinDeCelis2013}. Setting the observer-end
flux to zero adds a source-free monopole. With outward force defined as
positive, the leading self-force becomes
\[
 -\frac{q^2R}{2A^3}\left(1-\frac{2M}{R}\right)
 +\frac{q^2R}{2A^3}=\frac{q^2M}{A^3},
 \qquad A\gg R,
\]
the same leading repulsive self-force as a neutral black hole of mass $M$.
The far-field force's sign therefore does not determine whether a throat is
present and does not allow distinguishing a wormhole from a black hole in 
their setup.  Interestingly, equal asymptotic potentials need not always 
conflict with charge conservation:
in the asymptotically flat cylindrical wormhole, the observer-end flux is
independent of source position
\cite{RubinDeCelisSimeone2020}.

\FloatBarrier
\section{Constancy of the first-order mass monopole}
\label{sec:gravity}

At first order about Schwarzschild, the observer-side mass monopole is constant
even for nonspherical source motion: higher multipoles do not enter its vacuum
equations.

Take the observer-side background to be Schwarzschild,
$d\bar s^2=-f\,dt^2+f^{-1}dr^2+r^2d\Omega^2$, with
$f=1-r_g/r$.
The vacuum $\ell=0$ sector contains only a mass shift and a coordinate
change \cite{MartelPoisson2005SchwarzschildPerturbations}.
Removing the coordinate part, write the candidate mass perturbation as
\begin{equation}
 h_{tt}=\frac{2b(t)}{r},\qquad
 h_{rr}=\frac{2b(t)r}{(r-r_g)^2}.
 \label{eq:ds-observer-mode}
\end{equation}
These components are $h_{\alpha\beta}=b(t)\,\partial_M\bar g_{\alpha\beta}$,
with far-field acceleration shift $-b/r^2$.

\begin{samepage}
Replacing $f$ by $F(t,r)=1-2[M+b(t)]/r$ gives the linearized mixed
Einstein component
\begin{equation}
 \delta G^r{}_t=\frac{2\dot b}{r^2}.
 \label{eq:momentum-constraint}
\end{equation}
Vacuum therefore requires $\dot b=0$.  The constant mass shift is fixed
by the initial data, not by the instantaneous source position.
The argument is local to the observer's vacuum region; it assumes
neither vacuum at the throat nor any asymptotic structure.
\end{samepage}

In the DS construction, two Schwarzschild exteriors are joined at $R>r_g$.
Their revised preprint gives the observer-side perturbation
\eqref{eq:ds-observer-mode} with $b=\mu R/A$
\cite{DaiStojkovic2019ObservingV3}.
Letting $A$ follow an orbit gives $\dot b=-\mu R\dot A/A^2$, inconsistent
with the vacuum condition whenever $\dot A\ne0$.

DS allow mass shifts but fix equal clock rates at the two infinities;
their static matching then determines $b(A)$
\cite[Eqs.~(30)--(35)]{DaiStojkovic2019ObservingV3}.
Source motion may, however, change the relative clock rates between the ends
while the mass monopole remains fixed, paralleling the source-dependent
potential difference in the Maxwell analogue. A weak-field one-ended
wormhole provides an explicit example of
nonzero circulation of gravitational acceleration
\cite[Sec.~V.3]{FrolovKrtousZelnikov2023RingWormholes}.

Higher multipoles can still perturb a star's orbit through the throat.
At first order about Schwarzschild, the dipole contains only a locally
pure-gauge displacement and a constant angular-momentum shift, so the leading
time-dependent physical vacuum perturbation is quadrupolar
\cite{MartelPoisson2005SchwarzschildPerturbations,Thorne1980Multipole}.

The quasistatic quadrupolar acceleration falls as $r^{-4}$, compared with
$r^{-2}$ for the proposed DS field. Taking comparable field strengths at the
mouth scale $R$, the quadrupolar effect is therefore suppressed by order
$(R/r)^2$ at the observer. Using the approximate DS mouth scale
$R\sim0.1\,\mathrm{AU}$ and the star S2's
semimajor-axis distance $r\sim10^3\,\mathrm{AU}$
\cite[Sec.~V]{DaiStojkovic2019ObservingV3}, the quadrupolar effect is
roughly eight orders of magnitude smaller. Resonant excitation of long-lived
wormhole modes \cite{CardosoFranzinPani2016Ringdown} lies outside this
quasistatic estimate.

\section{Nonincrease of the observer-end mass}
\label{sec:nonlinear}

Beyond first order, gravitational radiation can carry energy away, but the
Bondi balance law still prohibits an increase in observer-end mass.

For a two-ended wormhole, assume regular Bondi asymptotics and vacuum near
the observer end's future null infinity. Its Bondi energy $E_B(u)$ obeys
\cite{BondiVanderBurgMetzner1962Gravitational,Sachs1962Asymptotic,
WaldZoupas2000Conserved}
\begin{equation}
 E_B(u_2)-E_B(u_1)
 =-\frac{1}{32\pi}\int_{u_1}^{u_2}du
    \int_{S^2}N_{AB}N^{AB}\,d\Omega\le0,
 \qquad u_2>u_1.
 \label{eq:bondi-integrated}
\end{equation}
Here $u$ is Bondi retarded time and $N_{AB}$ the news tensor. The law remains
valid during nonlinear evolution of the source and throat and requires no
energy condition on the supporting matter. DS's prescribed Schwarzschild
mass shift extends throughout the observer exterior, giving
$E_B=M+\mu R/A(u)$
\cite[Eqs.~(32)--(35)]{DaiStojkovic2019ObservingV3}.
The proposed periodic mass variation therefore violates the balance law.
Source motion may still modulate radiative mass loss.

When both mouths share one exterior, the Bondi energy belongs to the whole
system. To use the two-ended result, start with initial data whose two ends
are vacuum near infinity and have future-timelike ADM energy--momentum.
A gluing theorem
\cite[Theorem~1 and Corollary~1.1]{ChruscielCorvinoIsenberg2011NBody}
connects their distant parts through a common vacuum exterior while
preserving the data near the mouths and throat. Choose the preserved region
to contain all data influencing a fixed observer region and time interval,
including through either mouth. Assuming well-posed causal evolution,
causal uniqueness gives the same observer-side geometry at sufficiently
large exterior separation \cite{Sbierski2015CauchyDevelopment}.
The preserved region can grow without bound as the separation increases.
The two-ended Bondi law therefore excludes the proposed periodic mass signal
in this limit.

\section{Conclusion}
\label{sec:conclusion}

Matter moving near one mouth of a wormhole can affect stellar orbits near the
other. In vacuum, however, the mass inferred near that mouth remains fixed at
first order about Schwarzschild. The DS interpretation of a family of static
solutions with different monopoles as successive states of one orbiting system
is incompatible with the vacuum Einstein equations.
The constraint in the Maxwell case is exact:
charge measured around a mouth changes only when current crosses the
measuring surface.  Beyond first order, the observer-end Bondi mass can
decrease through radiation but cannot increase when the end is vacuum near
future null infinity. The proposed periodic mass signal is also absent in a
large-separation limit of wormholes with both mouths in one exterior.

Higher multipoles do reach the observer through the throat. At S2's semimajor-axis distance,
our quasistatic estimate gives a quadrupolar acceleration roughly eight orders
of magnitude below the proposed DS signal, taking comparable field strengths
at the mouth scale. Source motion may also modulate
radiative mass loss and, when both mouths share one exterior, directly influence
the observer through ordinary space.

\noindent\begin{minipage}{\textwidth}
\vspace{1.5\baselineskip}
\begin{acknowledgments}
The author thanks Ilia Slobodov for helpful discussions.

The author used OpenAI GPT-6.1 and Anthropic
Claude Opus~5.5 for scientific discussion and critique, literature searches,
editing, and figure preparation. The author
directed the exchanges, checked the arguments, calculations, references, and
figures, and takes full responsibility for the manuscript.
\end{acknowledgments}
\end{minipage}

\bibliographystyle{apsrev4-2}
\bibliography{references}

\end{document}